\documentclass{edm_template}
\usepackage{subfig}
\usepackage{hyperref}

\begin{document}

\title{Mining Frequent Learning Pathways from a Large Educational Dataset}

\numberofauthors{3}
\author{
Nirmal Patel \\ \affaddr{Playpower Labs} \\ \email{nirmal@playpowerlabs.com} \and
Collin Sellman \\ \affaddr{Arizona State University} \\ \email{collin.sellman@asu.edu} \and
Derek Lomas \\ \affaddr{Playpower Labs} \\ \email{derek@playpowerlabs.com}
}

\maketitle
\begin{abstract}
In this paper, we describe data mining techniques used to extract frequent learning pathways from a large educational dataset. These pathways were extracted as a directed graph that encoded student learning processes. Our dataset contains more than 800 million interactions of over 3 million anonymized students in an online learning platform. Performing process mining on large and complex datasets regularly yields incomprehensible process models. Although, if we cluster data and obtain groups following similar processes, we can greatly improve process mining results. To this end, we developed a sequence clustering algorithm that let us group students who followed similar learning pathways. To extract frequent learning pathways from these clusters of data, we developed a graph-based process discovery algorithm that revealed to us the sequences of learning activities that many students followed. These sequences represented \textit{highways} of student learning. 
\end{abstract}


\section{Introduction}
Digital learning platforms collect a wide variety of student interaction data. These data can be used to inform continuous improvement at the level of the product, school district, classroom or the individual student. Recently, we've begun to investigate individual sequences of instructional content, which we call "learning pathways" or "student journeys."  Using various data mining techniques, we've analyzed the learning pathways from thousands of students, revealing the emergence of interesting patterns and structures.

Association rule mining has been used to find interesting patterns within student learning logs \cite{garcia2011association}. For example, we can find association rules like \textit{If students fail in quiz X, what instructional content do they access afterward?} Often times, we find many interesting rules, but these rules are not linked and they do not tell us anything about entire learning pathways of students. By using newly developed process mining techniques, we can look for process models of student learning \cite{mukala2015exploring, trcka2010process}. Such models can be used to discover usage patterns or to compare usage to ideal patterns. Unfortunately, when working with data from many thousands of students, we find that there are too many underlying processes. Therefore, the resulting process models show "spaghetti": an unintelligible mess of various processes. Process discovery algorithms like Fuzzy Mining can help reduce the spaghetti \cite{gunther2007fuzzy}, but they render models that are not amenable to extensive manipulation or enhancement.

To address these issues, we first explored sequence clustering techniques to see if they could help us identify groups of students who follow similar processes \cite{hompesdiscovering}. We developed an edit distance based clustering algorithm, which allowed us to group together students with similar learning pathways. We found sequence clustering crucial for further analysis, as it significantly reduced the complexity of our process models. Using our clustering technique, we discovered three distinct groups of students that differed in their length of learning paths.

To discover process models within the clusters, we implemented a graph-based process discovery algorithm. This algorithm uses an iterative procedure to find out how frequently students transition from one activity to another in their learning paths. Applying this algorithm to our data yielded a directed graph. After visualizing the graph and removing infrequent edges from it, we found an intelligible process model. Further pruning of the graph showed us very clearly what learning pathways students followed. We found graph visualization to be a good tool for exploring student activity sequences. It allowed us to quickly evaluate our hypotheses about how students interacted with the learning platform. We also noted that our graph-based process models were easy to manipulate.

The rest of the paper is structured as follows. In section 2, we describe the dataset which we used for data mining. In section 3, we briefly describe the sequence clustering algorithm we used on data and discuss the results. In section 4, we succinctly describe the graph mining algorithm and show some examples from our analysis. At the end, we discuss the usefulness of techniques presented in this paper, and directions for future work.

\section{Dataset}
Our original dataset is formatted as an event log. Each row of the dataset corresponds to an event that captures learner interaction with an online learning platform. Examples of events are logging in, opening an activity, saving progress in an activity, submitting an activity etc. Millions of students across the United States interact with the platform. Within the platform, there are many programs available for different subjects and grade levels. We chose to analyze a grade 3 math program. This program is a digital curriculum divided into topics, lessons, and activities, and can be accessed via a web browser. There are different types of activities that students can do, including games, formative assessments, and summative assessments. Students can either do activities assigned by the teacher or choose to do activities they like. Teachers can also create their own assessments and upload them on the platform. For our analysis, we selected a particular formative assessment activity and randomly sampled 20,000 users from the population of students who access it. We chose sample size so as to keep our computations efficient.

The dataset contains many different variables, most of which are listed in Table 1. Roughly, these variables can be put into two major categories: organizational and instructional. Organizational variables tell us about the student's learning context and instructional variables tell about the student's interaction with digital assets and the platform. Since we were first interested in finding out what sequences of things students follow, we concentrated on instructional variables. For the formative assessment that we selected for analysis, multiple event types are possible. We selected the submitting event, which indicated that student had finished the assessment. Typically, this event also contains the student's score. We transformed the data and made them amenable to analysis by filtering the dataset and selecting particular variables. Specifically, we chose to extract three variables from our database: \textit{Student ID}, \textit{Activity ID} and \textit{Timestamp}. This way, we had access to the sequence of activities (in our case, formative assessments) that every student went through. These activities spanned across the entire grade 3 math curriculum. After we had extracted data in the above format and sampled the desired number of students, we used a sequence clustering algorithm to group together students with similar activity sequences.

\begin{table}
\centering
\caption{Variables of the dataset}
\begin{tabular}{ c c c } \hline
Organizational & Instructional & Other \\ \hline
Teacher ID  & Activity ID & Student ID \\
Class ID & Activity Type & Session ID \\
School ID & Event Type & Timestamp \\
District ID & Skill & \\
State & Score & \\
\hline\end{tabular}
\end{table}

\section{Clustering}
Exploration of our large dataset revealed to us that students rarely followed same learning paths. This fact posed a big challenge to our process mining work since high variance in student activity sequences kept giving us "spaghetti" process models. We hypothesized that by clustering together students who followed similar learning paths, we could get more intelligible process models. In the process mining paradigm, various algorithms have been proposed to cluster sequential data \cite{bose2009context}. Algorithms specific to clustering student activity sequences have also been proposed \cite{klingler2016temporally}. We developed and used an edit distance based clustering algorithm to cluster students with similar learning paths.

Edit distance (or more specifically, Levenshtein distance) between two strings is measured as the minimum number of characters we have to add, subtract and substitute to turn one string into another. We extended this idea to arbitrary alphabets, so the distance between two activity sequences was measured as the number of activities we had to add, subtract and substitute to turn one activity sequence into another. For example, distance between activity sequences \textit{<a,b,b,c,d,e,e>} and \textit{<a,b,c,d,e>} is 2, because by removing two activities from first sequence, we can turn it into second sequence in the least number of operations. We used this generalized distance metric to compute a distance matrix having pairwise distances between all student activity sequences. Several clustering methods are available to cluster distance matrices. Partitioning around medoids offers a partitioning approach but requires the number of clusters in advance. On the other hand, hierarchical clustering does not require the number of clusters in advance, so we used it to cluster our distance matrix. We used Ward's method for agglomeration. Hierarchical clustering can be done using various agglomeration criteria. We found that using Ward's method as a criterion gave us clusters where observations within clusters had similar paths and path lengths. Although other methods like average linkage could give clusters with higher within cluster path similarity, they suffered from an excessive amount of nesting. In contrast, Ward's method was able to produce more equally sized clusters.

Our results indicated the presence of 3 clusters which had meaningful differences. Properties of these clusters are listed in Table 2. Looking at the results, we inferred that Cluster 1 corresponded to students who did just a few items and stopped. Subsequent clusters appeared to be related to students with medium and high amounts of usage in the program. Cluster 3 students generally did more activities than cluster 2 students, i.e. they went further in using their digital curriculum.

\begin{table}
\centering
\caption{Clustering results}
\begin{tabular}{ c | c c c } \hline
                     & Cluster 1 & Cluster 2 & Cluster 3 \\ \hline
Avg. sequence length & 2.80       & 13.72      & 44.85      \\
SD sequence length   & 1.91      & 5.79       & 18.91      \\
Number of students   & 10524     & 5524      & 3952      \\
\hline \end{tabular}
\end{table}

\section{Graph Mining}
To mine a graph that encoded student learning pathways, we followed a simple procedure. We started by creating an $n \times n$ zero matrix $\mathcal{M}$ where $n$ is the number of unique activities that students did. Entry $\mathcal{M}_{ij}$ corresponded to how many times students did activities $i$ and $j$ in sequence. We iterated through data of all students, and for every pair of subsequent activities that students did, we added 1 to the appropriate cell of $\mathcal{M}$. At the end of the procedure, we produced an adjacency matrix of a directed graph.

Every node in this graph corresponded to an activity that a student could do, and edges of the graph showed pathways that students had taken. Edge weights of the graph represented how many times that edge was traveled by students. We ran the graph mining algorithm on clusters we had discovered earlier. Table 3 compares the number of nodes and edges of graphs corresponding to clusters.

\begin{table}
\centering
\caption{Graph mining results}
\begin{tabular}{ c | c c c } \hline
                    & Cluster 1 & Cluster 2 & Cluster 3 \\ \hline
Nodes               & 123       & 169       & 474       \\
Edges               & 3277      & 5378      & 6662      \\
\hline \end{tabular}
\end{table}

We noted that even though the number of students decreased from Cluster 1 to 3, the number of nodes and edges increased. We chose to focus on Cluster 3, where students learning pathways were substantially longer than in other clusters and the graph was denser. Figure 1 (a) shows the learning pathway graph of Cluster 3, without any edge filtering and changes in node size. We can see that it is impossible to make any inferences visually (it looks like "spaghetti"). Even after clustering, infrequent paths visually obscured mainstream student behavior. To see frequent paths, we removed infrequent edges in the graph. Figure 1 (b) shows a graph where edges with less weight are filtered out and node sizes are proportional to node degrees. Now we see highways of student learning, i.e. paths that many students took. When we filtered the graph in Figure 1 (b) further, we saw even fewer paths, and these paths could be investigated directly by looking at the visualization. Figure 2 shows those few paths. Figure 3 zooms into a mid left region of Figure 2 to show three activities whose paths are frequently traversed. We confirmed that these activities follow each other in the digital curriculum too.

\begin{figure}
\centering
\subfloat[Unfiltered graph]{{\includegraphics[scale=.19]{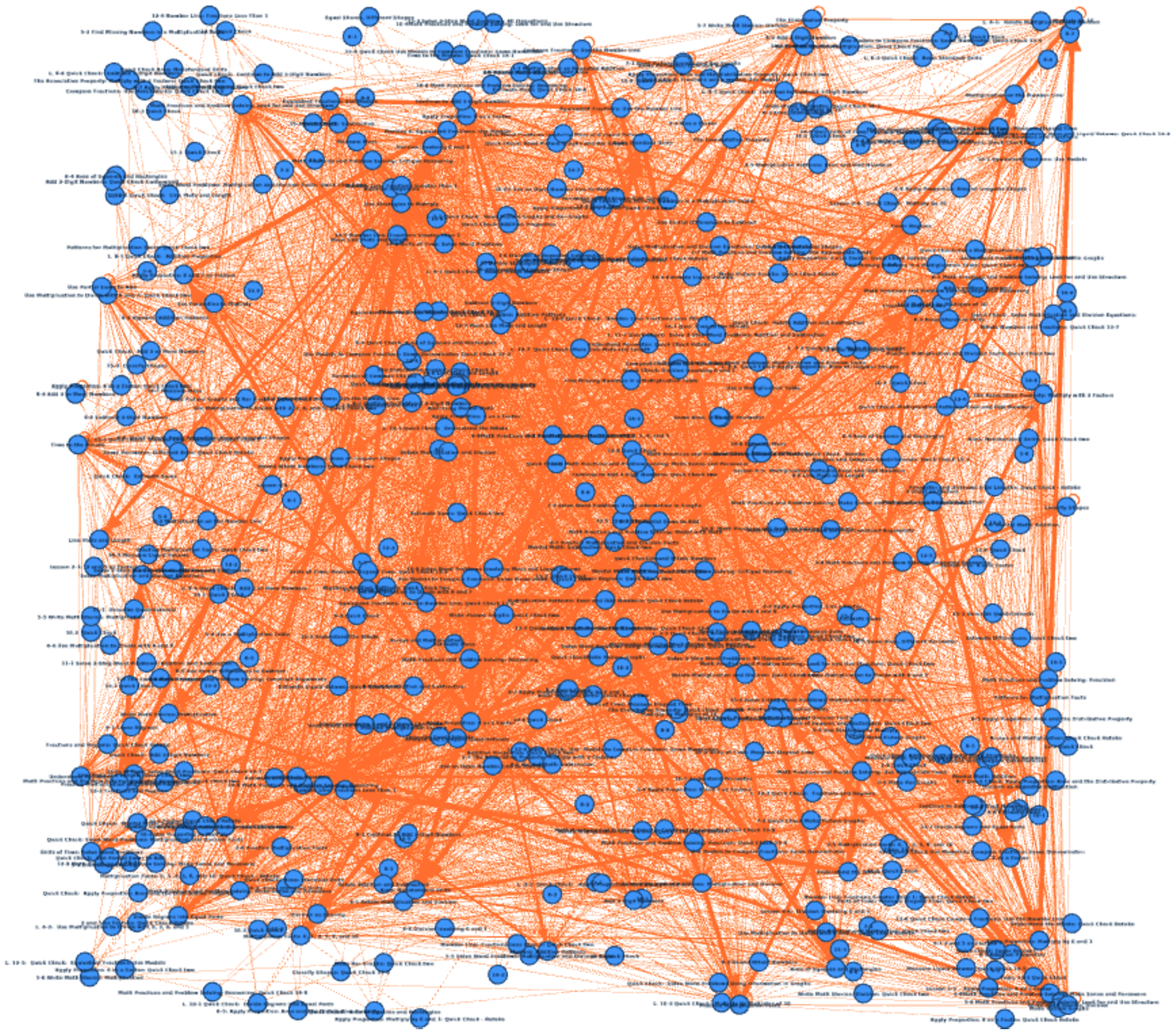} }}%
\qquad
\subfloat[Filtered graph]{{\includegraphics[scale=.19]{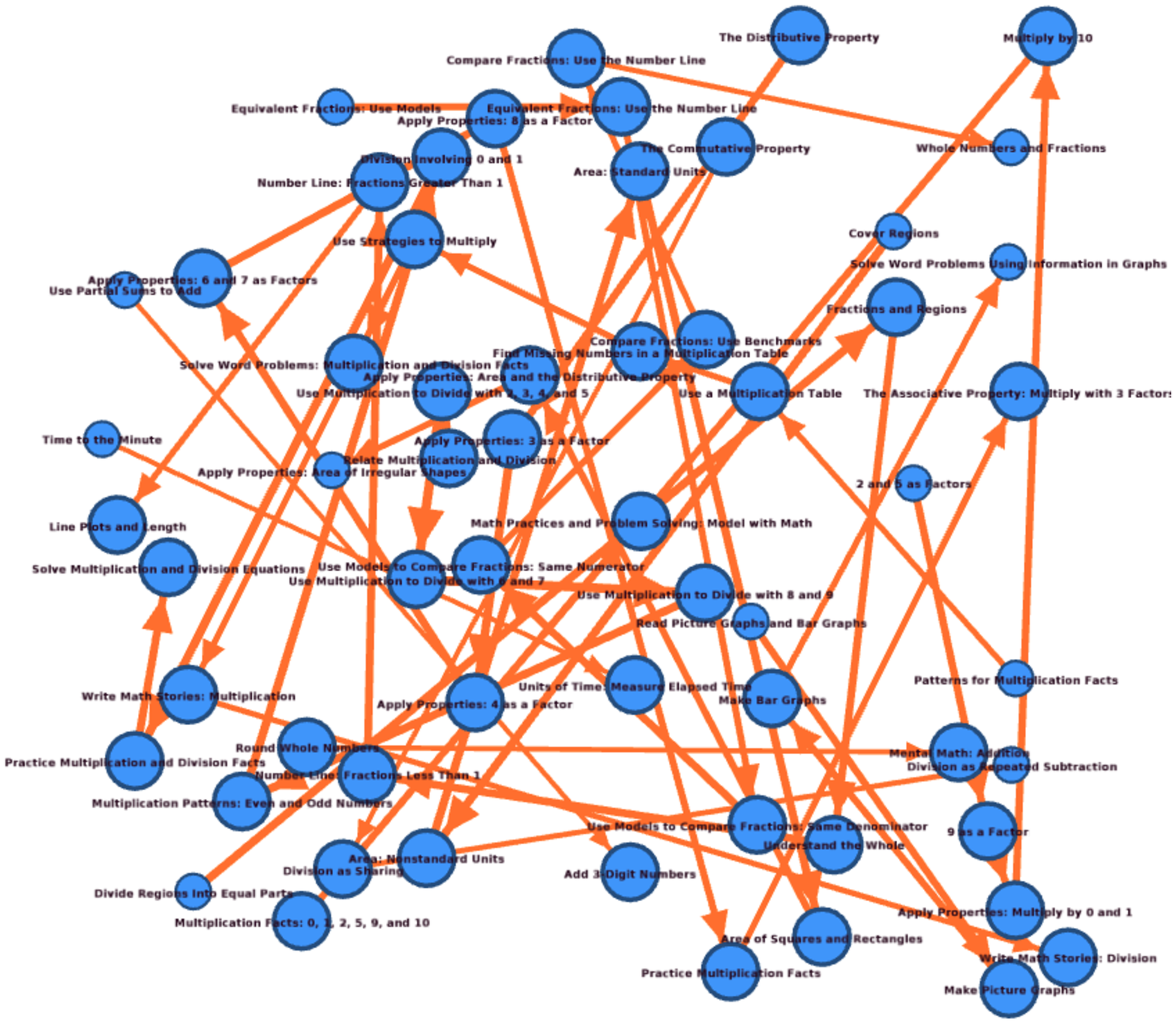} }}%
\caption{Learning pathway graphs}
\end{figure}

\begin{figure}
\centering
\includegraphics[scale=.38]{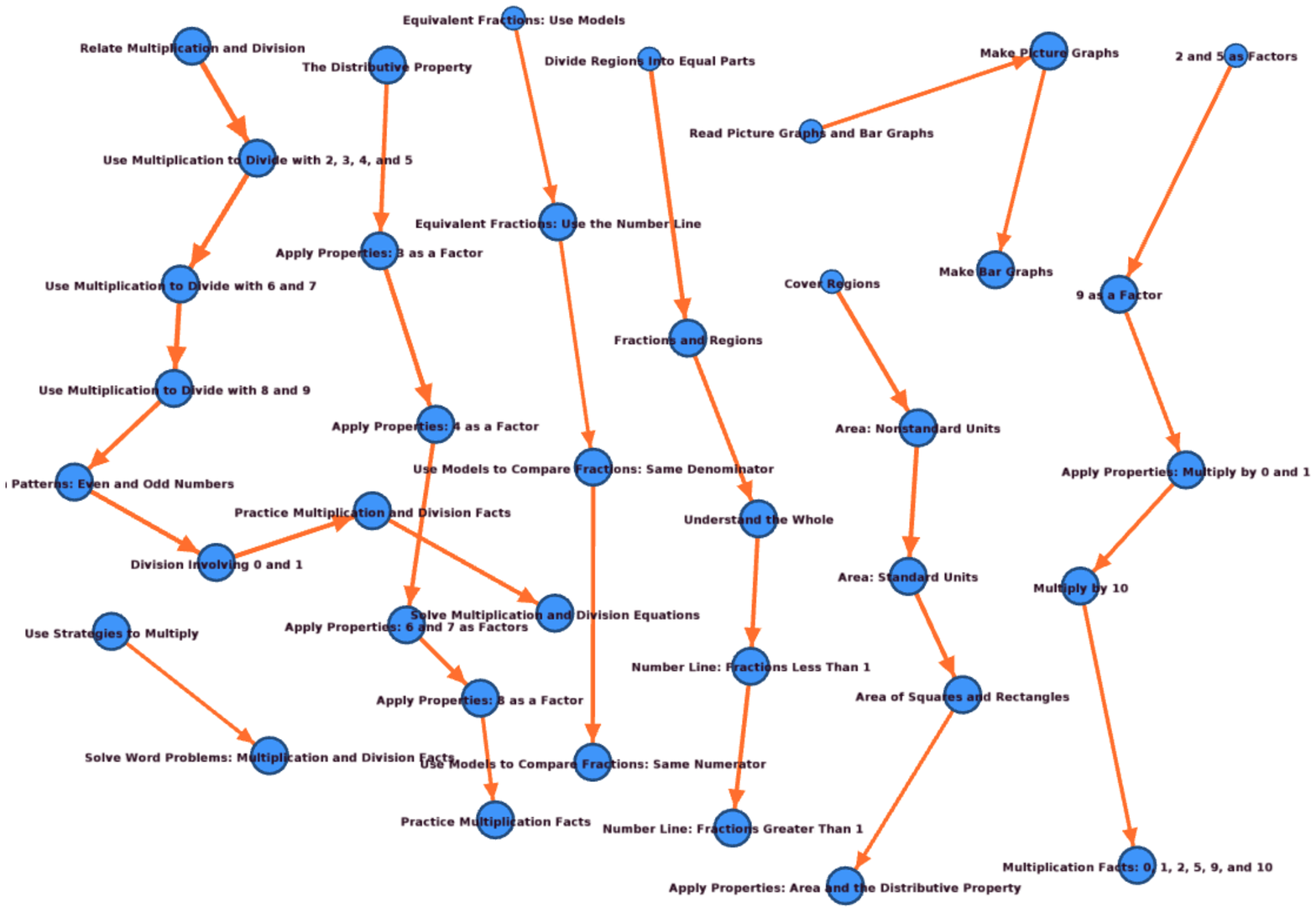}
\caption{Some highways of student learning}
\end{figure}

\begin{figure}
\centering
\includegraphics[scale=.45]{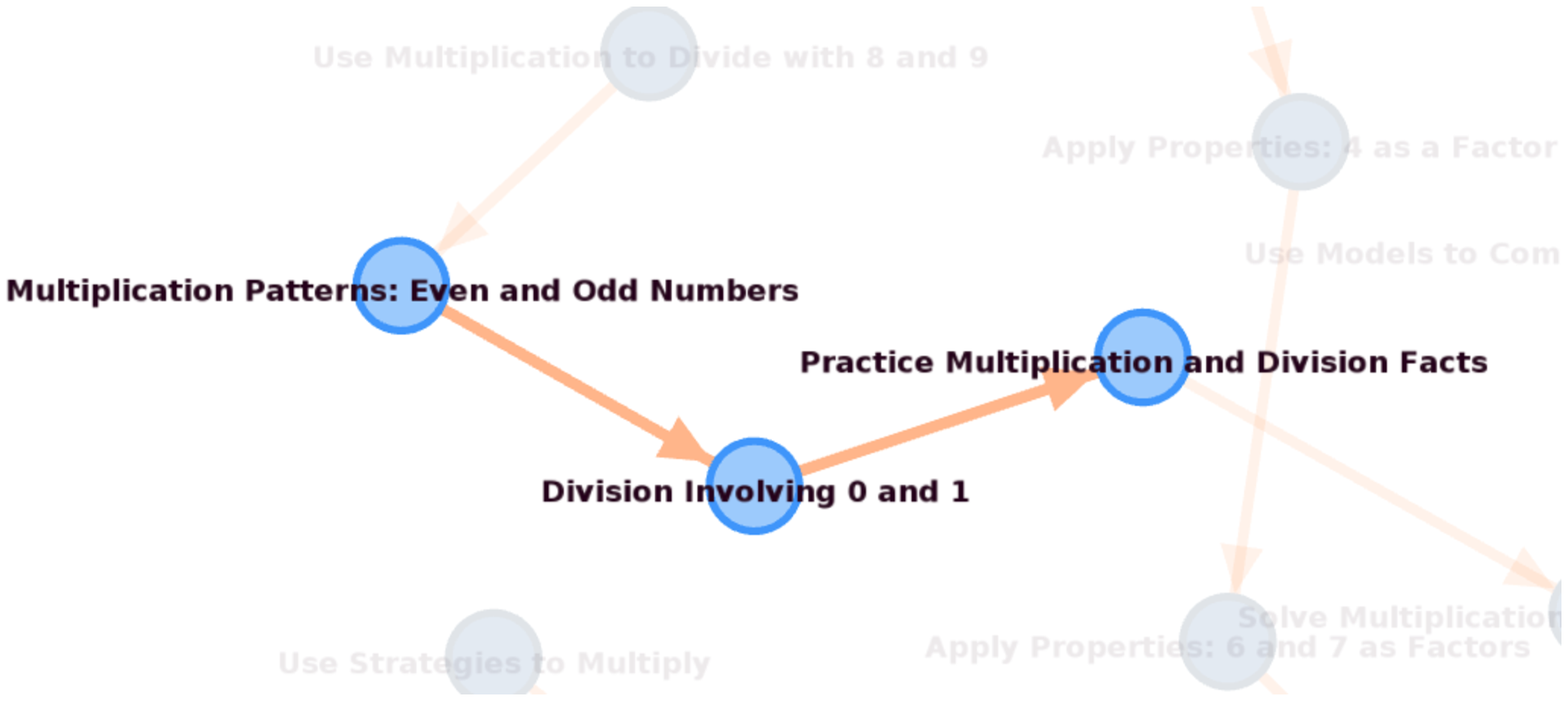}
\caption{Three activities that follow each other many times}
\end{figure}

\section{Conclusions}
By performing sequence clustering and graph-based process mining on educational data, we were able to identify student groups with similar usage patterns and examine their learning pathways. While we applied our technique to a smaller subset of data, we believe that it is a generalizable technique that can be applied to any educational dataset from which we can extract student activity sequences. We can analyze student learning pathway graphs to explore how different paths affect student and class performance. Teachers and curriculum designers can learn the sequence of activities their students generally go through during their learning and whether these sequence of activities conform to a desired or designed sequence. Last but not least, designers of digital learning products can get insights into user behavior that can help them make informed design decisions towards data-driven continuous improvement. We believe the methods presented in this paper can provide information that can be of use to various stakeholders involved in education, by showing them what students do and creating more accurate models of student behavior in digital environments. Open source \texttt{R} implementation of sequence clustering technique presented in this paper is available online\footnote{\url{https://github.com/nirmalpatel/learning_pathway_mining}}.

\section{Future Work}
Additional student data related to their activity and performance (such as scores) can easily be added to student pathways. We are currently mining decision points in graphs that will explore how performance on one activity influences the choice of the next activity which is a rich area for future research. This is similar to decision mining in process models \cite{van2016process}. Since we have skills mapped to each activity in our dataset, we also hope to explore the skill acquisition processes of students. In the methods domain, computational challenges related to the scalability of our procedures also remain when using larger, unsampled datasets. We believe these can be addressed by using a parallel computing framework like Apache Spark.

\bibliographystyle{abbrv}
\bibliography{main}

\begin{thebibliography}{1}

\bibitem{bose2009context}
R.~J.~C. Bose and W.~M. van~der Aalst.
\newblock Context aware trace clustering: Towards improving process mining
  results.
\newblock In {\em Proceedings of the 2009 SIAM International Conference on Data
  Mining}, pages 401--412, 2009.

\bibitem{garcia2011association}
E.~Garc{\'\i}a, C.~Romero, S.~Ventura, C.~de~Castro, and T.~Calders.
\newblock Association rule mining in learning management systems.
\newblock {\em Handbook of educational data mining}, pages 93--106, 2011.

\bibitem{gunther2007fuzzy}
C.~W. G{\"u}nther and W.~M. Van Der~Aalst.
\newblock Fuzzy mining--adaptive process simplification based on
  multi-perspective metrics.
\newblock In {\em International Conference on Business Process Management},
  pages 328--343. Springer, 2007.

\bibitem{hompesdiscovering}
B.~Hompes, J.~Buijs, W.~van~der Aalst, P.~Dixit, and J.~Buurman.
\newblock Discovering deviating cases and process variants using trace
  clustering.
\newblock In {\em Proceedings of the 27th Benelux Conference on Artificial
  Intelligence}, pages 5--6, 2015.

\bibitem{klingler2016temporally}
S.~Klingler, T.~K{\"a}ser, B.~Solenthaler, and M.~Gross.
\newblock Temporally coherent clustering of student data.
\newblock In {\em Proceedings of the 9th International Conference on
  Educational Data Mining, International Educational Data Mining Society},
  pages 102--109, 2016.

\bibitem{mukala2015exploring}
P.~Mukala, J.~Buijs, and W.~Van~der Aalst.
\newblock Exploring students' learning behaviour in moocs using process mining
  techniques.
\newblock Technical report, BPMcenter.org, 2015.

\bibitem{trcka2010process}
N.~Trcka, M.~Pechenizkiy, and W.~Van~der Aalst.
\newblock Process mining from educational data.
\newblock {\em Handbook of Educational Data Mining}, pages 123--142, 2011.

\bibitem{van2016process}
W.~M. van~der Aalst.
\newblock {\em Process mining: data science in action}.
\newblock Springer, 2016.
\newblock pg. 294.

\end{thebibliography}

\end{document}